\documentclass[conference]{IEEEtran}
\IEEEoverridecommandlockouts
\usepackage{amsmath,amssymb,amsfonts}
\usepackage{array}
\usepackage{booktabs}
\usepackage{cite}
\usepackage{graphicx}
\usepackage{multirow}
\usepackage{subcaption}
\usepackage{textcomp}
\usepackage{xcolor}
\usepackage{listings}
\usepackage{pgfplots}
\usepackage{amssymb,mathtools}
\usepackage{comment}
\usepackage[hidelinks]{hyperref}
\usepackage{algorithm}
\usepackage[noend]{algpseudocode}
\usepackage[para,flushleft]{threeparttable}
\usepackage[normalem]{ulem}
\usepackage{bm}

\algnewcommand{\LineComment}[1]{\State \(\triangleright\) #1}
\useunder{\uline}{\ul}{}

\definecolor{codegreen}{rgb}{0,0.6,0}
\definecolor{codegray}{rgb}{0.5,0.5,0.5}
\definecolor{codepurple}{rgb}{0.58,0,0.82}
\definecolor{backcolour}{rgb}{0.95,0.95,0.92}

\lstdefinestyle{mystyle}{
    backgroundcolor=\color{backcolour},   
    commentstyle=\color{codegreen},
    keywordstyle=\color{magenta},
    numberstyle=\tiny\color{codegray},
    stringstyle=\color{codepurple},
    basicstyle=\ttfamily\footnotesize,
    breakatwhitespace=false,         
    breaklines=true,                 
    captionpos=b,                    
    keepspaces=true,                                  
    showspaces=false,                
    showstringspaces=false,
    showtabs=false,                  
    tabsize=2,
    frame=single,
}

\begin{document}

\title{
    SATAY: A Streaming Architecture Toolflow for Accelerating YOLO Models on FPGA Devices
}

\author{
     \IEEEauthorblockN{
         Alexander Montgomerie-Corcoran, 
         Petros Toupas, 
         Zhewen Yu and 
         Christos-Savvas Bouganis
     }
    \IEEEauthorblockA{
    Imperial College London, UK \\
    \{
        alexander.montgomerie-corcoran15, 
        petros.toupas21,
        zhewen.yu18, 
        christos-savvas.bouganis
    \}@imperial.ac.uk}
}

\maketitle

\begin{abstract}
AI has led to significant advancements in computer vision and image processing tasks, enabling a wide range of applications in real-life scenarios, from autonomous vehicles to medical imaging. 
Many of those applications require efficient object detection algorithms and complementary real-time, low latency hardware to perform inference of these algorithms.
The YOLO family of models is considered the most efficient for object detection, having only a single model pass. 
Despite this, the complexity and size of YOLO models can be too computationally demanding for current edge-based platforms. 
To address this, we present SATAY: a Streaming Architecture Toolflow for Accelerating YOLO.
This work tackles the challenges of deploying state-of-the-art object detection models onto FPGA devices for ultra-low latency applications, enabling real-time, edge-based object detection. 
We employ a streaming architecture design for our YOLO accelerators, implementing the complete model on-chip in a deeply pipelined fashion.
These accelerators are generated using an automated toolflow, and can target a range of suitable FPGA devices.
We introduce novel hardware components to support the operations of YOLO models in a dataflow manner, 
and off-chip memory buffering to address the limited on-chip memory resources.
Our toolflow is able to generate accelerator designs which demonstrate competitive performance and energy characteristics to GPU devices, and which outperform current state-of-the-art FPGA accelerators.
\end{abstract}

\section{Introduction}
\label{sec:intro}
Object detection is a fundamental concept in computer vision, enabling machines to identify and locate objects in images or videos. It has wide-ranging applications, from autonomous driving \cite{feng2021review} to surveillance systems \cite{wiangtong2023deployment} and medical imaging \cite{jaeger2020retina}. Advancements in object detection, driven by novel model designs, aim to achieve real-time performance.  One category of such extremely fast and efficient models is the YOLO family, which achieves state-of-the-art performance in object detection benchmarks\cite{everingham2010pascal, lin2014microsoft}.

Accelerating YOLO networks on FPGA devices can significantly improve their inference speed, enabling faster execution compared to traditional CPU or GPU implementations, and are well-suited for real-time object detection. Many previous research works \cite{preusser2018inference, kathail2020xilinx} focused on the optimisation of the computationally intensive convolutional layers on FPGA fabrics, and offloaded the other operations to CPUs. Such heterogeneous overlay-based designs offer flexibility to deploy various versions of YOLO networks;
however, they are severely constrained by the data communication bottlenecks between FPGAs and CPUs.

Other works \cite{nguyen2020layer, li2021mapping, herrmann2022yolo}, have explored the deployment of the entire YOLO network onto the FPGA device, by building customised computation units and data flows into their accelerator designs. As such, the impact of data communication bottlenecks is minimised and the overall performance is enhanced. However, the proposed accelerator designs in these works are tailored to specific versions of the YOLO network, and lack the versatility to target more modern object detection models. The lack of easy adaptation hinders the hardware community from keeping up with the latest advancements in the state-of-the-art versions of the YOLO family.

\begin{figure}[!t]
    \centering
    \includegraphics[width=1.0\columnwidth]{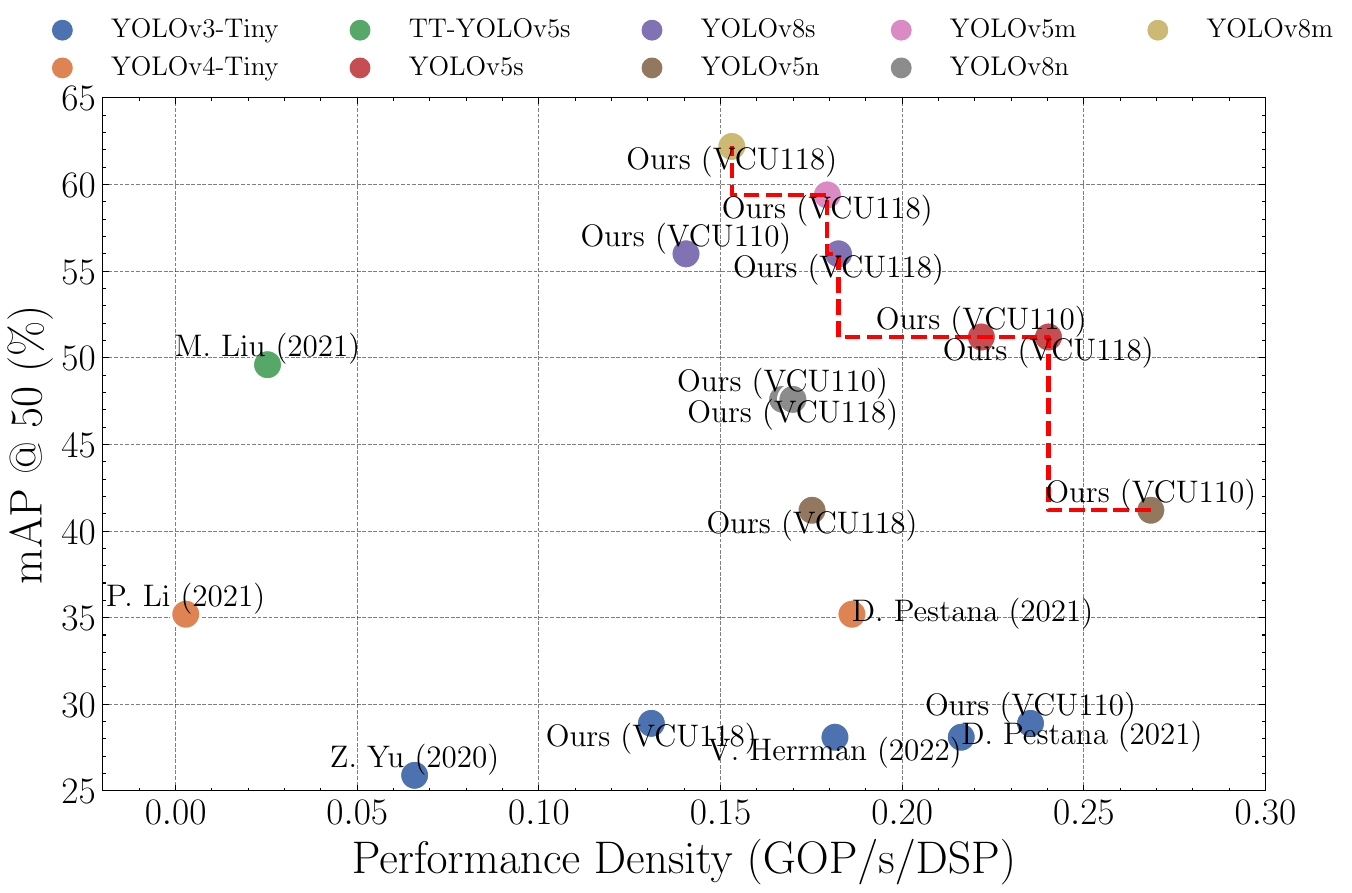}
    \caption{
        Comparison between the accuracy and performance of FPGA accelerator designs generated by our SATAY toolflow and prior works, for the COCO dataset.
        Our toolflow demonstrates state-of-the-art performance for YOLO model inference, defining the Pareto frontier.
    }
    \label{fig:acc-perf-pareto}
\end{figure}

In this paper, we introduce a toolflow that enables the mapping of diverse models from the YOLO family onto a wide range of FPGA devices. Our toolflow is designed to achieve both high performance and versatility by tackling the following challenges:

\paragraph{Diverse Operation Libraries}

When accelerating YOLO networks on FPGAs, it is crucial to have efficient hardware implementations for both standard operations, such as convolution and max pooling, as well as custom operations specific to YOLO models, such as leaky ReLU, SiLU, and resize. 

\paragraph{Long Skip Connections}

YOLO networks rely on multi-scale feature fusion to detect objects of different sizes. 
This process involves merging feature maps from different stages of the network, which leads to long skip connections in the computation graph. 
This requires extensive buffer placement for streaming architectures.

\paragraph{Large Design Space}

Over the generations of YOLO models, there has been a discernible trend towards deeper network architectures, with 47 new layers introduced since YOLOv3.
The increased complexity and size of the network architectures has brought increased difficulty in finding optimal hardware configurations.

Our toolflow addresses these challenges by introducing novel streaming hardware components for YOLO-specific operations,
applying fine-grain quantization methods to alleviate on-chip memory resource constraints,
and implementing a \textit{software FIFO} to further reduce the on-chip memory usage.
Automated design space exploration is performed with these, generating hardware designs which surpass edge-GPU performance, and achieve state-of-the-art performance compared to FPGA accelerator designs.

\section{Background \& Related Work}

\subsection{Object Detection}

Object detection is a fundamental computer vision task that aims to identify and locate objects within images or videos. This task involves detecting the presence of objects in an image and accurately determining their spatial boundaries, while assigning a specific category or label to each detected object.

To achieve accurate object detection, state-of-the-art methods can be broadly classified into two categories:

\paragraph*{One-stage Methods}
such as YOLO (You Only Look Once)\cite{redmon2016yolo}, SSD (Single Shot MultiBox Detector)\cite{wei2016ssd}, and RetinaNet\cite{thung-yi2017retinanet}, prioritise fast inference speed. These models efficiently detect objects by directly predicting bounding boxes and class probabilities in a single pass through the network.

\paragraph*{Two-stage Methods}
including Faster R-CNN (Region Convolutional Neural Network)\cite{shaoqing2017fasterrcnn}, Mask R-CNN (Mask Region Convolutional Neural Network)\cite{kaiming2017maskrcnn}, and Cascade R-CNN\cite{zhaowei2018cascade}, focus on achieving higher detection accuracy. These approaches typically involve a preliminary region proposal step followed by a fine-grained object classification and localisation step. Although two-stage methods are computationally more expensive, they often yield superior detection performance.

\subsection{The YOLO Model Family}

Since its initial release \cite{redmon2016yolo} in 2016, several versions of YOLO have been developed, each improving on accuracy and efficiency. 
The major limitation of the initial YOLO model was its inability to detect small objects.
In order to enhance the network's ability to detect objects of various sizes, YOLOv2 \cite{redmon2017yolov2}, v3 \cite{redmon2018yolov3} and v4 \cite{bochkovskiy2020yolov4} incorporated anchor boxes, feature pyramid networks, and path aggregation networks in their designs respectively.

The basic building components of the network have also witnessed improvements in computational efficiency. From YOLOv4, the plain feedforward convolutions have been replaced by the Cross Stage Partial (CSP) Blocks, which is a variation of the residual block that incorporates cross-stage connections for enhanced feature representation \cite{bochkovskiy2020yolov4}. In YOLOv6 and v7, the Reparameterized Block has been incorporated which includes convolutions with different kernel sizes in parallel at training time \cite{li2022yolov6, li2023yolov6, wang2022yolov7}.

The advancements in the YOLO family have enabled accurate and efficient object detection. However, they have also increased the number of operation types and the complexity of layer interconnection. This presents research opportunities in hardware accelerator designs on FPGA devices, aiming to outperform CPUs and GPUs.
In this work, we focus on the YOLOv3, YOLOv5 and YOLOv8 families of architectures, as they represent the most dramatic changes in network architecture as well as accuracy and workload, across the generations of models.

\begin{figure*}[!ht]
    \centering
    \includegraphics[width=0.85\textwidth]{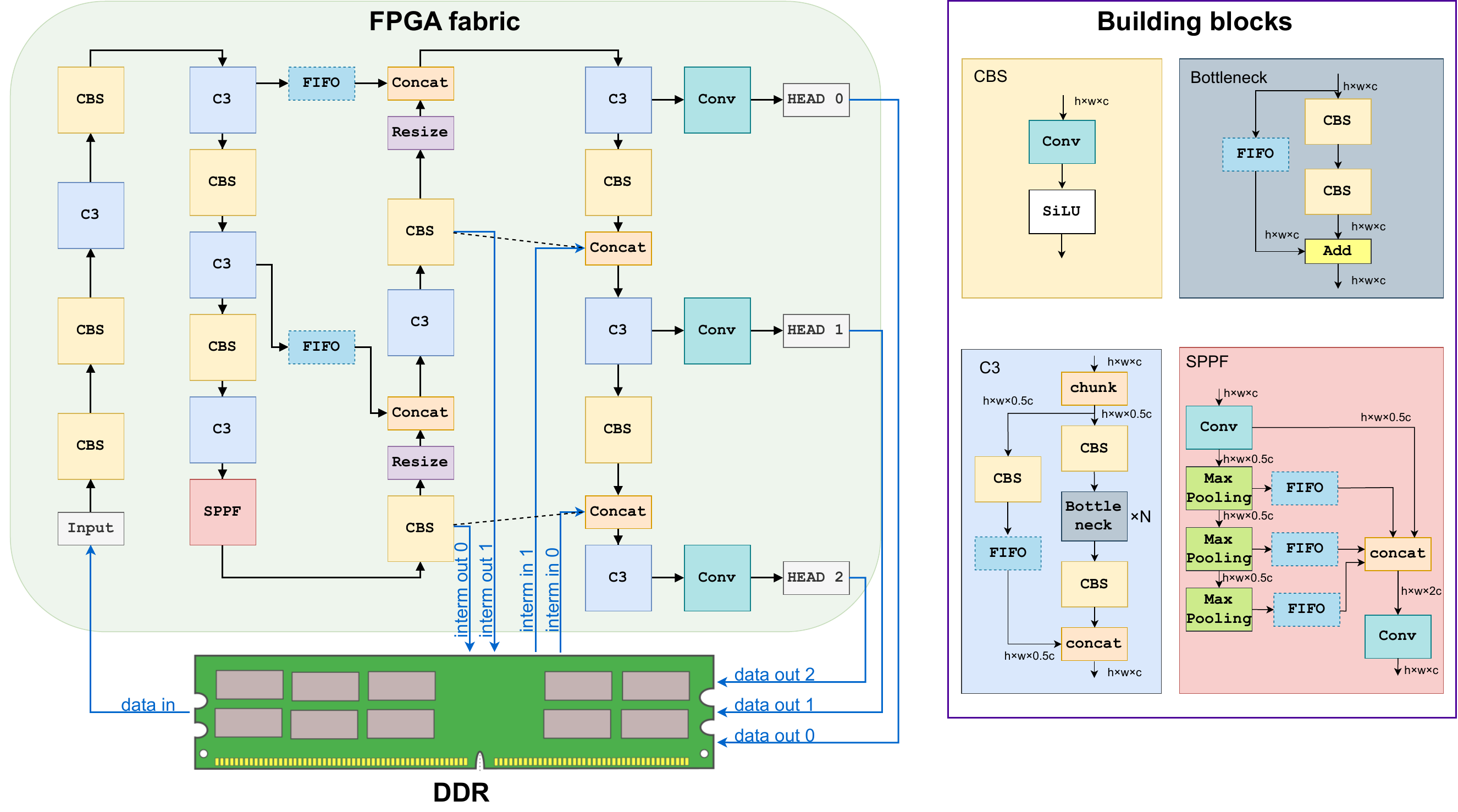}
    \caption{
        An example of the YOLOv5 accelerator generated by our toolflow. 
        \textbf{Black} arrows show on-chip data flow, while \textcolor{blue}{\textbf{blue}} arrows show off-chip data flow. 
        \textbf{Dashed black} lines show the skip connections that are diverted to off-chip (Section \ref{subsec:sw_fifo}).
    }
    \label{fig:YOLOv5n_accelerator}
\end{figure*}

\subsection{FPGA Object Detection Acceleration}

The field of research into object detection acceleration has focused on one-stage methods \cite{hongxiang2018ssdlite, Zhao2017OptimizingCO}, due to their efficient network architectures.
Several studies have explored FPGA-based accelerators for YOLO networks in particular, capitalising on the flexibility and customisation offered by FPGAs. 
Accelerators for YOLO-v2 were implemented in \cite{nakahara2018lightweight, preusser2018inference}, with notable performance advantages over embedded CPUs and GPUs. 
Both studies employed weight binarization techniques to reduce computation costs and memory requirements. 
Additionally, they adopted a sequential computation order, allowing the convolutional layers to share the computation resource effectively. Building upon this sequential computation approach, \cite{nguyen2020layer}, \cite{li2021mapping} and \cite{liu2021efficient} further extend this line of work to YOLOv3, v4 and v5 respectively, accommodating the customised operations involved in each version. As network designs became more compact and less redundant in evolving versions, the quantization strategies in these studies also became less aggressive. \cite{nguyen2020layer} utilised a combination of 1-bit and 8-bit quantization, while both \cite{li2021mapping} and \cite{liu2021efficient} opted for 16-bit quantization instead.

The aforementioned studies focused on designing accelerators for specific versions of the YOLO family, achieving notable computational efficiency. 
However, there is a considerable time delay between advancements in machine learning and hardware communities. 
Currently, there is a lack of published performance results for running the latest YOLOv8 on FPGA devices owing to the challenges discussed in Section~\ref{sec:intro}.
In this work, we present a toolflow that enables the mapping of various YOLO models (from v3 to v8) onto a wide range of FPGA devices. 
Our approach differs from related work such as Vitis-AI \cite{kathail2020xilinx}, which builds heterogeneous accelerators across CPUs and FPGAs. 
Instead, we focus solely on FPGA devices and employ a fully pipelined streaming architecture to maximise performance and efficiency.

\section{Hardware Architecture}

This section outlines the streaming architecture and the hardware building blocks which have been designed and tailored to the YOLO model family.

\subsection{Streaming Architecture}

A streaming architecture \cite{venieris2016fpgaconvnet, harflow3d, stan2022hpipe, blott2018finn} design principle is taken for the FPGA accelerators generated by our toolflow.
Streaming architectures follow an elastic circuit \cite{elastic-circuits} design paradigm, where computation nodes communicate in a pure data flow manner, with ready/valid handshake interfacing.
In the case of neural network acceleration, these computation nodes correspond to nodes in the machine learning graph.
This design style leads to deeply pipelined architectures that make efficient use of on-chip routing bandwidth.

To illustrate the general data flow of a generated design, Figure~\ref{fig:YOLOv5n_accelerator} shows an example block diagram of a YOLOv5 accelerator."
It can be seen that each machine learning operation is implemented as a dedicated hardware block within the FPGA fabric.
The execution of the hardware begins when the input image is streamed into the hardware from off-chip DDR memory, with an `NHWC" format.
The computational pipeline within the FPGA processes the input image, producing the bounding box predictions at the output, which are returned to off-chip memory.
The only other external connections are that of the off-chip buffer, which is discussed in Section~\ref{subsec:sw_fifo}.
During execution, all the model's parameters stay on-chip, requiring no off-chip memory accesses.

The streaming architecture approach taken has many benefits over alternative single-engine accelerators.
The designs are able to exploit the inherent parallelism found in YOLO model architectures.
Storing the weights on-chip dramatically reduces the off-chip memory bandwidth requirements, a bottleneck commonly seen in most other architecture designs.

\subsection{Building Blocks}
\label{subsec:hw_lib}

The streaming architectures described in this work are composed of a number of building blocks.
The design of such building blocks follows a dataflow approach, with ready/valid interfaces,
and are tailored to the set of operations required by the YOLO family, in particular the \textit{Resize} and \textit{HardSwish} components. 
Parallelism can be exploited in all of the building blocks and it is discussed in Section~\ref{subsec:dse}.

\begin{figure}[h!]
    \centering
    \includegraphics[width=0.80\columnwidth]{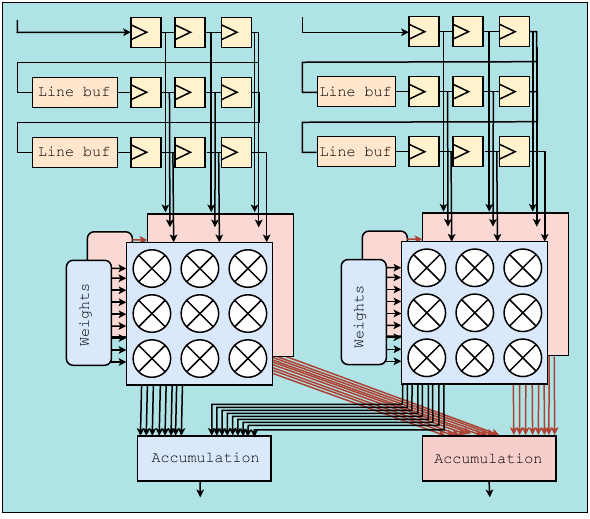}
    \caption{
    Convolution building block.
    }
    \label{fig:convolution}
\end{figure}

\paragraph{Convolution}

The hardware implementation of the \textit{convolution} operation is shown in Figure~\ref{fig:convolution}.
A sliding window generator comprised of line buffers feeds a matrix-vector multiplication engine, generating partial sums which are then accumulated.
The parallelism across inputs as well as within the kernel itself is illustrated.
The sliding window generator leads to very efficient usage of on-chip memory, requiring only $(K-1) \times W \times C$ of words of a feature map to be stored on-chip. 
The matrix-vector multiplication engine contains an array of $K \times K$ DSPs.
Parameters for the convolution are stored on-chip in URAM and BRAM, alleviating the memory bandwidth bottleneck common in systolic array architectures.

\begin{figure}[h!]
    \centering
    \includegraphics[width=0.7\columnwidth]{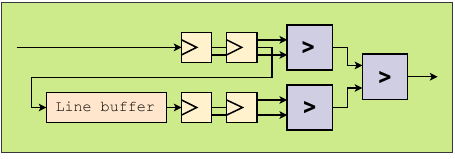}
    \caption{
    Max pooling building block.
    }
    \label{fig:max-pooling}
\end{figure}

\paragraph{Max Pooling}
The \textit{max pooling} operation utilises a similar architecture to the convolution hardware.
A sliding window generator feeds a comparator tree, which discovers the maximum value in the window.
This component also benefits from the efficient on-chip memory needed by the sliding window generator.

\begin{figure}[h!]
    \centering
    \includegraphics[width=0.5\columnwidth]{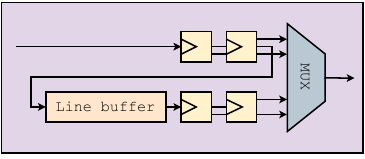}
    \caption{
    Resize building block.
    }
    \label{fig:resize}
\end{figure}

\paragraph{Resize}

The \textit{resize} hardware uses a sliding window generator to cache a window of words from the current and previous rows of the feature map.
The words are then selected and duplicated by a \texttt{MUX} with data-dependent control logic.
This novel hardware performs the resizing operation on the fly, requiring minimal buffering.

\begin{figure}[h!] 
    \begin{subfigure}{0.3\columnwidth}
        \centering
        \includegraphics[width=\linewidth]{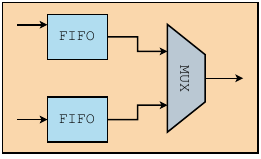}
        \caption{Concat}
        \label{fig:concat}
    \end{subfigure} 
    \hfill
    \begin{subfigure}{0.285\columnwidth}
        \centering
        \includegraphics[width=\linewidth]{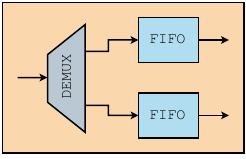}
        \caption{Split}
        \label{fig:split}
    \end{subfigure} 
    \hfill
    \begin{subfigure}{0.225\columnwidth}
        \centering
        \includegraphics[width=\linewidth]{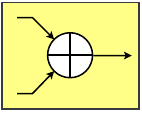}
        \caption{Add}
        \label{fig:add}
    \end{subfigure}
    \caption{
        Branching and merging building blocks.
    }
    \label{fig:concat_split_add}
\end{figure}

\paragraph{Split, Concat and Add}

These components constitute the branching and merging operations seen in YOLO models.
The \textit{split} hardware separates channels of the feature map, which appear sequentially in the stream, using a demultiplexer.
Conversely, the \textit{concat} hardware combines the channel dimensions of several feature maps using a stream multiplexer.
Both components require buffering for the channel dimension to prevent unnecessary back-pressure.
The \textit{add} hardware performs an element-wise addition of data from two streams.

\begin{figure}[h!] 
    \begin{subfigure}{0.53\columnwidth}
        \centering
        \includegraphics[width=\linewidth]{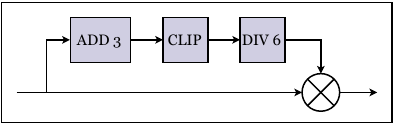}
        \caption{HardSwish}
        \label{fig:hardswish}
    \end{subfigure} 
    \hfill
    \begin{subfigure}{0.28\columnwidth}
        \centering
        \includegraphics[width=\linewidth]{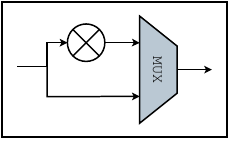}
        \caption{Leaky ReLU}
        \label{fig:leaky-relu}
    \end{subfigure} 
    \caption{
        Activation building blocks.
    }
    \label{fig:hardware}
    \vspace{-0.5cm}
\end{figure}

\paragraph{Leaky ReLU and HardSwish}
The activation functions used in the latest YOLO models are Leaky ReLU and SiLU, both of which are significantly more resource-intensive than ReLU.
In particular, the SiLU operation involves computing the sigmoid function, which requires costly floating point hardware.
Instead, the \textit{HardSwish} activation \cite{hardswish} is used ($x \cdot RELU6(x+3)/6$), which has a lot more efficient hardware implementation, as seen in Figure~\ref{fig:hardswish}.
Despite HardSwish being an approximation of the SiLU function, we demonstrate in Section~\ref{sec:eval} that it has a negligible impact on accuracy.
The implementation of \textit{Leaky ReLU} involves a multiplication with a constant, as well as 
a multiplexer for inputs above zero.

\section{Design Space Exploration}
This section outlines the design space exploration performed to discover efficient designs for YOLO accelerators.
The design space exploration outlined is a component of a toolflow that automates the mapping of a YOLO model to FPGA bitstream.
The stages of the toolflow are outlined below.
\begin{enumerate}
    \item \textbf{Parsing:} An ONNX model is parsed into an Internal Representation (IR).
    \item \textbf{DSE:} Design space exploration is performed on the IR.
    \item \textbf{Generation:} A bitstream is generated from the IR.
\end{enumerate}

This work uses the open-source toolflow SAMO~\cite{montgomerie2022samo} as a skeleton framework, which follows a similar approach to the other toolflows, namely fpgaConvNet \cite{venieris2016fpgaconvnet} and FINN \cite{blott2018finn}.
The design space exploration in this work focuses on resource allocation for both increasing performance and meeting on-chip resource constraints. 
For reference, a taxonomy of symbols used in this section is given in Table~\ref{tab:taxonomy}.

\begin{table}[!h]
\centering
\caption{Taxonomy of symbols used during DSE.}
\label{tab:taxonomy}
\begin{tabular}{@{}cl@{}}
    \toprule
    \textbf{Symbol} & \textbf{Description} \\
    \midrule
    $H, W$  & Spatial dimensions of feature map \\    
    $C$     & Channel dimension of feature map \\    
    $F$     & Filter dimension in convolution \\    
    $K$     & Kernel size of convolution and pooling \\    
    \midrule
    $w_w, w_a$  & Wordlength of weights and activations \\  
    \midrule
    $p_n \in \bm{p}$                    & Parallelism of node $n$ \\   
    $t_{n,m}^{buf} \in \bm{t}^{buf}$    & Type of buffer between nodes $n$ and $m$ \\   
    \bottomrule
\end{tabular}
\end{table}

\begin{figure*}[!htb]
    \begin{subfigure}{0.32\textwidth}
        \centering
        \includegraphics[width=\linewidth]{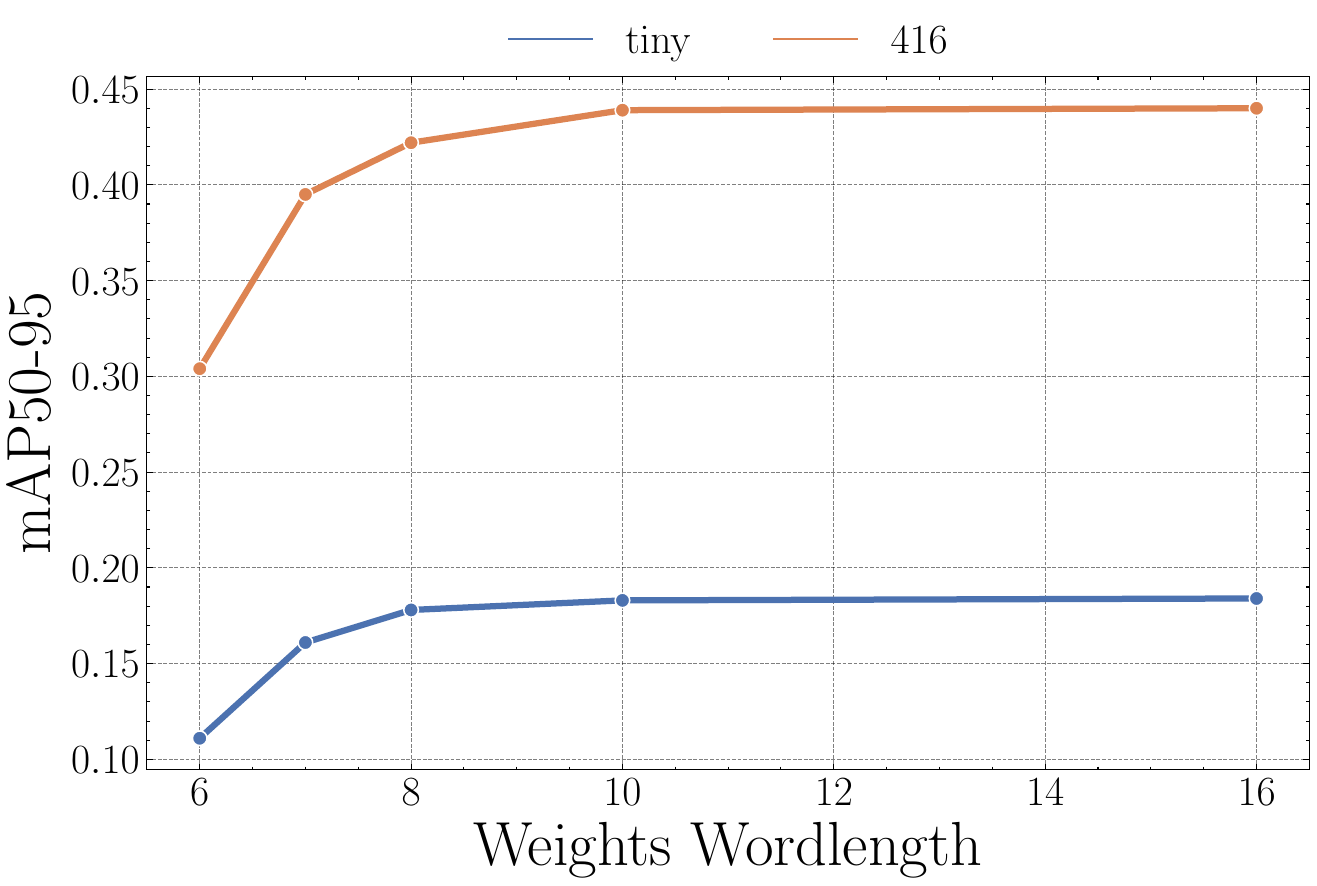}
        \caption{YOLOv3}
        \label{fig:yolov3_quant}
    \end{subfigure}
    \hfill %
    \begin{subfigure}{0.32\textwidth}
        \centering
        \includegraphics[width=\linewidth]{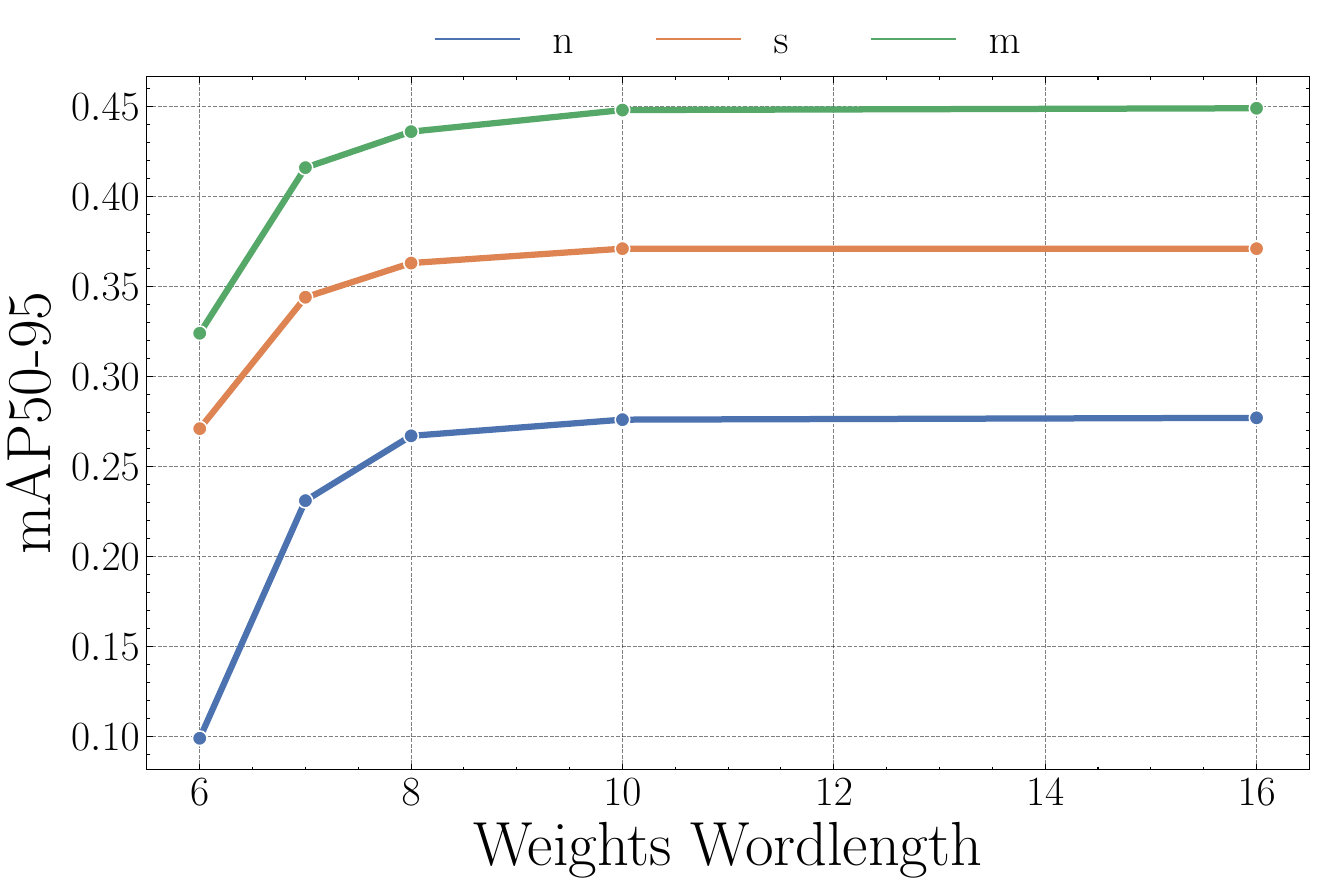}
        \caption{YOLOv5}
        \label{fig:yolov5_quant}
    \end{subfigure}
    \hfill %
    \begin{subfigure}{0.32\textwidth}
        \centering
        \includegraphics[width=\linewidth]{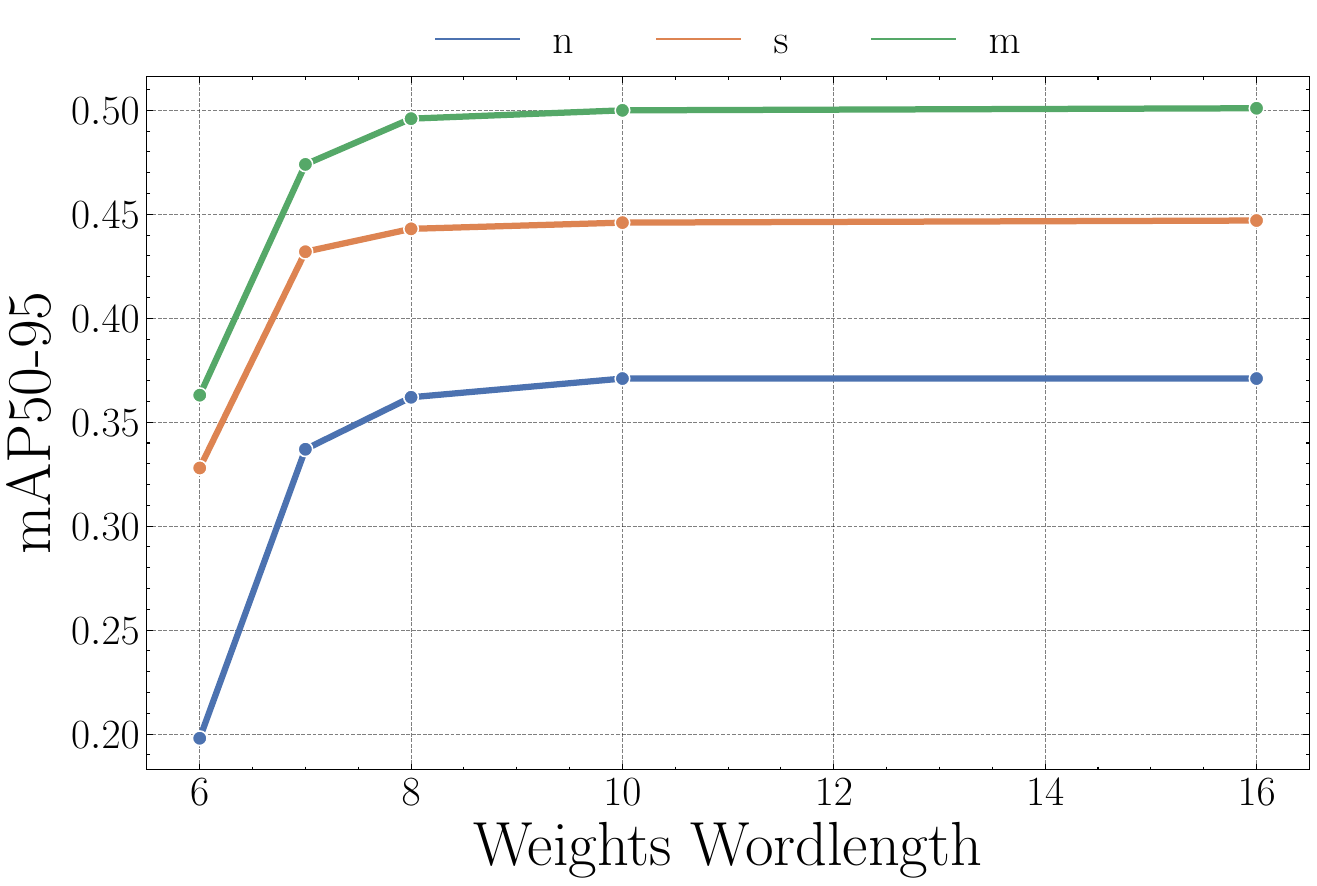}
        \caption{YOLOv8}
        \label{fig:yolov8_quant}
    \end{subfigure}
    \caption{Quantization results on YOLOv3, YOLOv5, and YOLOv8 and all of their variants. The x-axis shows the exploration of different weight wordlengths. The activation's wordlength is fixed at 16 bits throughout these experiments. Models are evaluated on the COCO-val2017 dataset.}
    \label{fig:yolo_models_quant}
    \vspace{-0.3cm}
\end{figure*}

\subsection{Quantization}

Due to the requirement of storing parameters on-chip, quantization techniques are employed to drastically reduce the resources required. 
During the parsing step, our toolflow accepts onnx models in the floating point format and simulates the quantization effect using the onnx runtime.
We adopt a layer-wise blocking floating point format, where the weights are quantized as follows:
\begin{equation}
    w' = round( \frac{w}{S} - Z) 
\end{equation}
$w$ is the original floating-point weights and $w'$ is its integer format. The parameters scale $S$ and zero point $Z$ are chosen based on the minimum and maximum of the weights within the same layer, as well as the wordlength $w_w$ chosen for quantization.
\begin{equation}
    S = \frac{w_{max} - w_{min}}{2^{L} - 1}
\end{equation}
\begin{equation}
    Z = round(w_{min} * S) + 2^{(L - 1)}
\end{equation}
Our quantization process is performed post-training, and directly applied to pre-trained weights without additional fine-tuning. We further explore the impact of wordlength choice for weights quantization, as depicted in Figure~\ref{fig:yolo_models_quant}. The results indicate that quantization has minimal effect on the accuracy of YOLO models, as long as the wordlength $w_w$ is at least 8 bits. In all these experiments, activations are also quantized, with their wordlength $w_a$ fixed at 16 bits.

\subsection{Performance Optimisation}
\label{subsec:dse}

Within streaming architectures, the performance is dictated by the slowest node within the accelerator.
The performance can be improved by increasing parallelism within the node, at the cost of increased resources. 
The resources that are affected by increased parallelism are the look-up table and embedded multiplier resources.
As can be seen in Table~\ref{tab:fpga-comparison}, the main constraint on designs is the DSPs, and therefore we focus our resource-constrained optimisation solely on DSP allocation.

\begin{algorithm}[h]
\caption{DSP Allocation Algorithm}
\begin{algorithmic}
\LineComment{Initialise all parallelism factors to 1}
\State $p_n = 1$ $\forall \; n \in [1..N]$;
\LineComment{Iterate until all DSPs are utilised}
\While {$\sum_{n=1}^N r^{DSP}(n) < \mathcal{R}^{DSP}$}
    \State $n = 1$ 
    \Comment{Start with the first node}
    \State $\mathcal{L}_{base} = \mathcal{L}(\bm{p})$ 
    \Comment{Calculate baseline latency}
    \State $\Delta_{prev} = 0$ 
    \Comment{zero performance improvement}
    \For{$m$ \textbf{in} $[1..N]$}
        \LineComment{Increase the parallelism of node $m$}
        \State $\Tilde{\bm{p}} = \{p_1, ..., p_{m-1}, p_m + 1, p_{m+1}, ..., p_N \}$
        \LineComment{Performance improvement of increasing $p_m$}
        \State $\Delta_{m}= \mathcal{L}_{base} - \mathcal{L}(\Tilde{\bm{p}})$ 
        \LineComment{Check if improvement greater than previous best}
        \If{ $\Delta_{m} < \Delta_{prev}$}
            \State $n = m$
            \Comment{Update best candidate node}
            \State $\Delta_{prev} = \Delta_{m}$
            \Comment{Update best performance delta}
        \EndIf
    \EndFor
    \State $p_n = p_n + 1$ 
    \Comment{Increase parallelism of node}
\EndWhile
\end{algorithmic}
\label{alg:dsp-allocation}
\end{algorithm}

To assist with the design space exploration, models of performance and resources are developed. 
The purpose of modelling is to allow for rapid evaluation of a design point and alleviates any need for performing synthesis and place and route in order to obtain performance and resource metrics.
This allows for magnitudes greater design points to be evaluated within a reasonable period of time.
To begin with, a model for the latency of a node $n$ in the accelerator is given as,
\begin{equation*}
    \label{eq:latency}
    l(n, \bm{p}) = \begin{cases}
        \frac{H_n \cdot W_n \cdot C_n \cdot F_n}{p_n \cdot f_{clk}}  & \text{if Convolution} \\
        \hfil \frac{H_n \cdot W_n \cdot C_n}{p_n \cdot f_{clk}}    & \text{otherwise}
    \end{cases}
\end{equation*}
where $l(n, \bm{p})$ is the latency model, $p_n$ is the parallelism factor, and $f_{clk}$ is the clock frequency of the design.
The general principle of the latency model is that each node has a workload to execute, which will take that many number of cycles.
By increasing parallelism within the node, this workload can be executed in parallel, reducing the latency by the parallelism factor.
The latency for the whole design with $N$ nodes is described as,
\begin{equation*}
    \mathcal{L}(\bm{p}) = \left( \max_{n \in [1..N]} l(n, \bm{p}) \right)+ \sum_{n=1}^{N} \frac{d(n)}{f_{clk}}
\end{equation*}
where $d(n)$ is the pipeline depth in cycles of each node $n$.
This latency model states that the latency is dictated by the slowest node in the accelerator plus the time taken for data to pass through the pipeline.

To evaluate the impact on resources, a model of DSP usage for each type of node is given below.
\begin{equation*}
    r^{DSP}(n, \bm{p}) = \begin{cases}
        K^2 \cdot p_n       & \text{if Convolution} \\
        \hfil 2 \cdot p_n   & \text{if HardSwish} \\ 
        \hfil p_n           & \text{if Leaky ReLU} \\ 
        \hfil 0             & \text{otherwise}
    \end{cases} 
\end{equation*}

Using the performance and resource models, an optimisation problem can be formulated for maximising the performance of the accelerator design, as shown below,
\begin{align*}
    \max_{\bm{p}} \, \mathcal{L}(\bm{p}) & \hfill \;\;\; \text{ s.t. } \sum_{n=1}^N r^{DSP}(n, \bm{p}) < \mathcal{R}^{DSP}
\end{align*}
where $\mathcal{R}^{DSP}$ is the available DSP resources of the FPGA. 
To solve this optimisation, a greedy algorithm was created, and is described in Algorithm~\ref{alg:dsp-allocation}.

\subsection{Memory Allocation}
\label{subsec:sw_fifo}

\begin{figure}[b!]
\centering
\begin{minipage}[c]{0.8\columnwidth}
\begin{lstlisting}[
columns=fullflexible,
frame=single,
language=python,
linewidth=\columnwidth,
basicstyle=\ttfamily\tiny,
caption={A simplified implementation of the software FIFO in Python, which can be used with the PYNQ framework.},
label={lst:soft-fifo},
]
def fifo(dma_in, dma_out, depth, chunk_size=256):

    # set the chunk counters
    chunk_cntr_in, chunk_cntr_out = 0, 0

    # allocate 2D array of memory
    mem = np.array((depth, chunk_size))

    # execute until transfer is complete
    while cntr_in < depth and cntr_out < depth:

        # transfer from PL to PS
        if cntr_in < depth:
            dam_in.receive(mem[cntr_in])
            cntr_in += 1

        # transfer from PS to PL
        if cntr_out < depth and cntr_out < cntr_in:
            dam_out.send(mem[cntr_out])
            cntr_out += 1
\end{lstlisting}
\end{minipage}
\vspace{-0.6cm}
\end{figure}

The memory allocated on-chip consists of three components: weight parameter storage, feature map caching in the sliding window line buffers, and buffering required for skip connections.
A summary of the relative size of these memory components alongside where in the system they are allocated is shown in Table~\ref{tab:mem-rsc}.

\begin{table}[!h]
    \centering
    \caption{Summary of typical memory types and utilisation.}
    \label{tab:mem-rsc}
    \begin{tabular}{@{}lcl@{}}
    \toprule
    \textbf{Memory Type} & \textbf{Utilisation} & \textbf{On-Chip} \\
    \midrule
    Weights                             & 50\% - 89\% & Yes \\
    Feature Maps (Sliding Window)       &  4\% - 20\% & Yes \\
    Feature Maps (Skip Connections)     &  7\% - 30\% & Partial \\
    \bottomrule
    \end{tabular}
\end{table}

The weights storage contributes to the majority of memory required by the streaming architecture system.
Due to the deeply pipelined hardware architecture, the weights are required to stay on-chip as their bandwidth requirements are too large to transfer off-chip.
This can lead to large on-chip memory requirements, with up to 26MB of memory needed within the FPGA.
Therefore there is motivation to reduce the on-chip resources of the next largest memory component, the skip connection buffers.

The deep connections between components of YOLO models require extensive buffering for a single partition streaming architecture.
Figure~\ref{fig:YOLOv5n_accelerator} illustrates the variety of skip connections used in YOLO models, and the corresponding FIFOs required. 
To reduce the on-chip memory resources required by implementing buffers on-chip, we instead move the buffering off-chip by implementing an off-chip software FIFO. 

\begin{figure*}[!htb]
    \begin{subfigure}{0.32\textwidth}
        \centering
        \includegraphics[width=\linewidth]{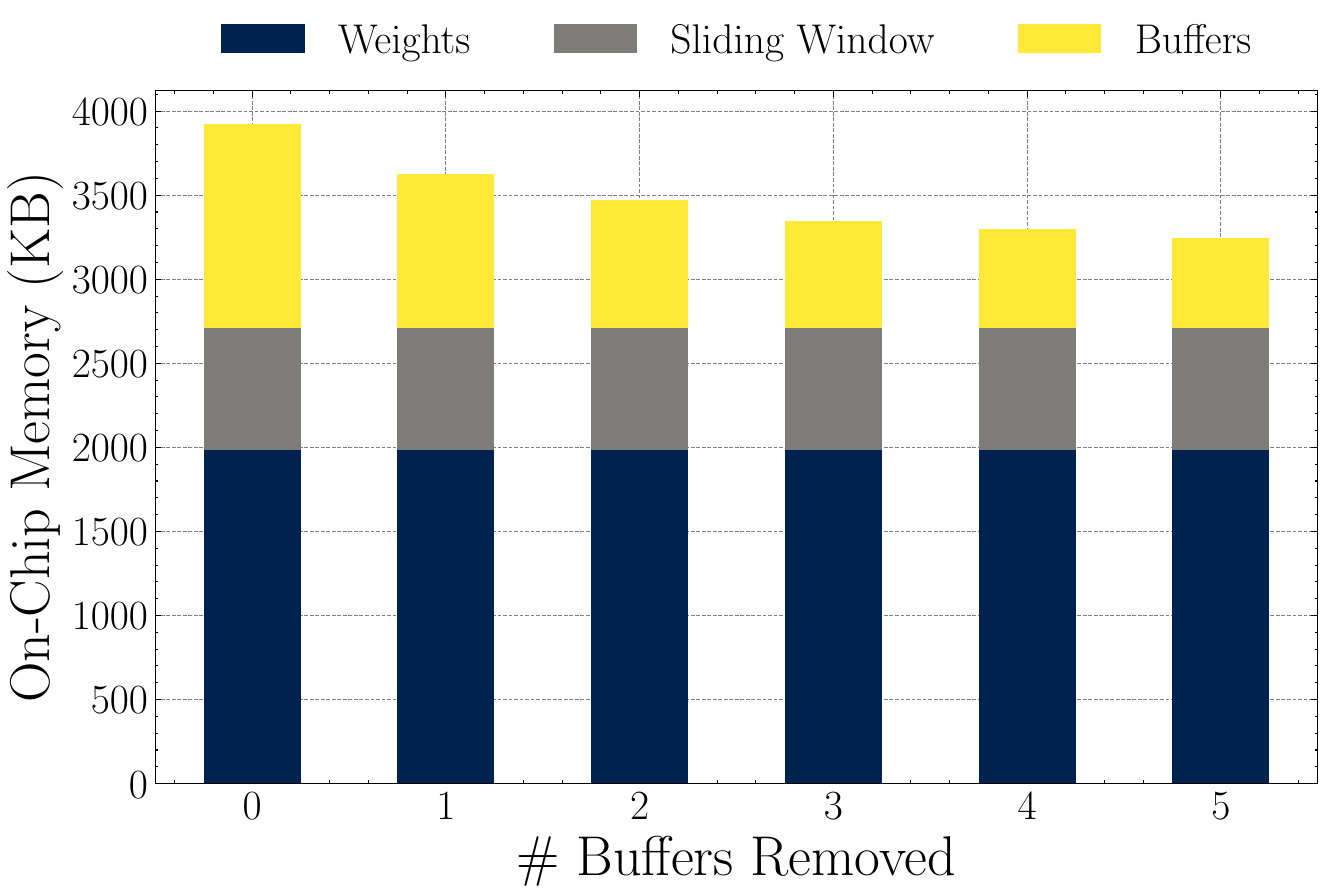}
        \caption{On-Chip Memory Resource Usage}
        \label{fig:ablation-buffer-size}
    \end{subfigure}
    \hfill %
    \begin{subfigure}{0.32\textwidth}
        \centering
        \includegraphics[width=\linewidth]{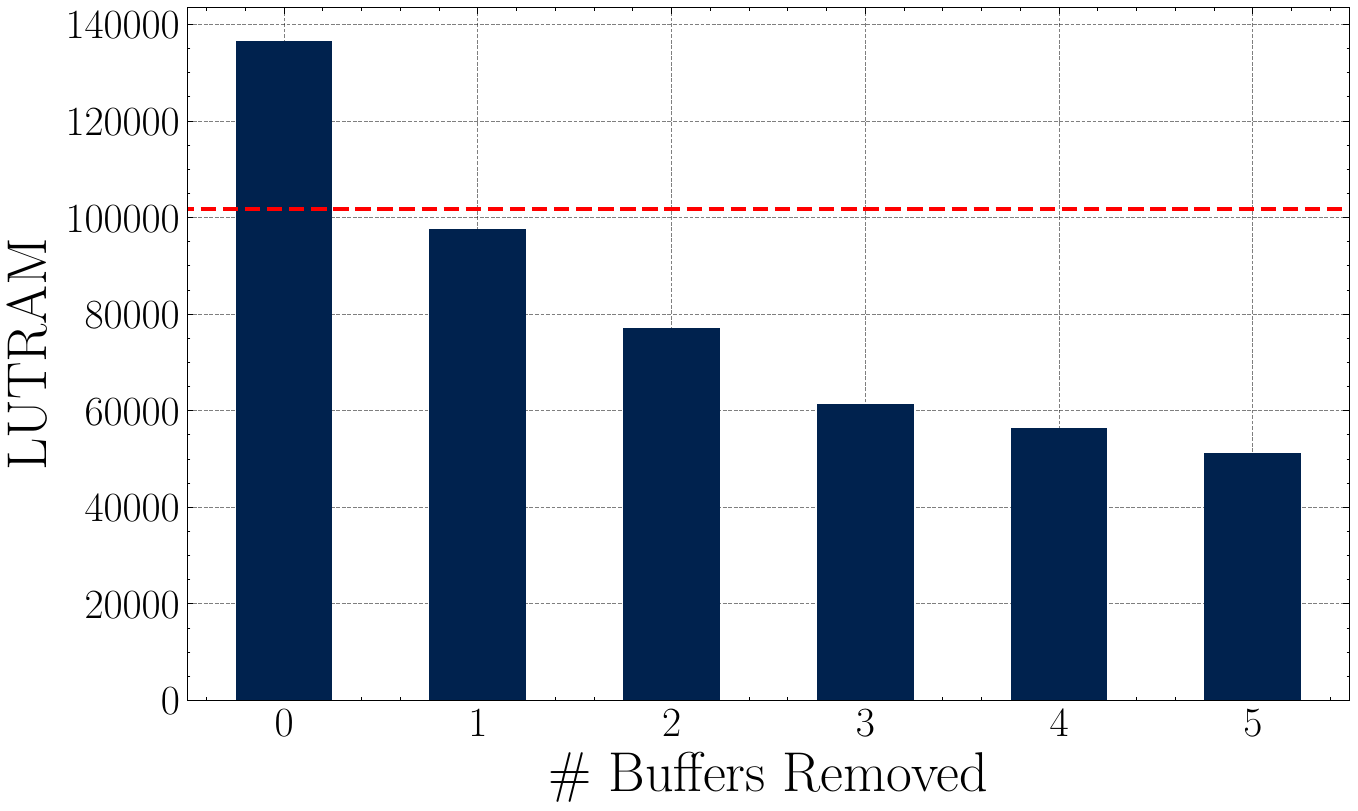}
        \caption{LUTRAM Usage}
        \label{fig:ablation-lutram}
    \end{subfigure}
    \hfill %
    \begin{subfigure}{0.32\textwidth}
        \centering
        \includegraphics[width=\linewidth]{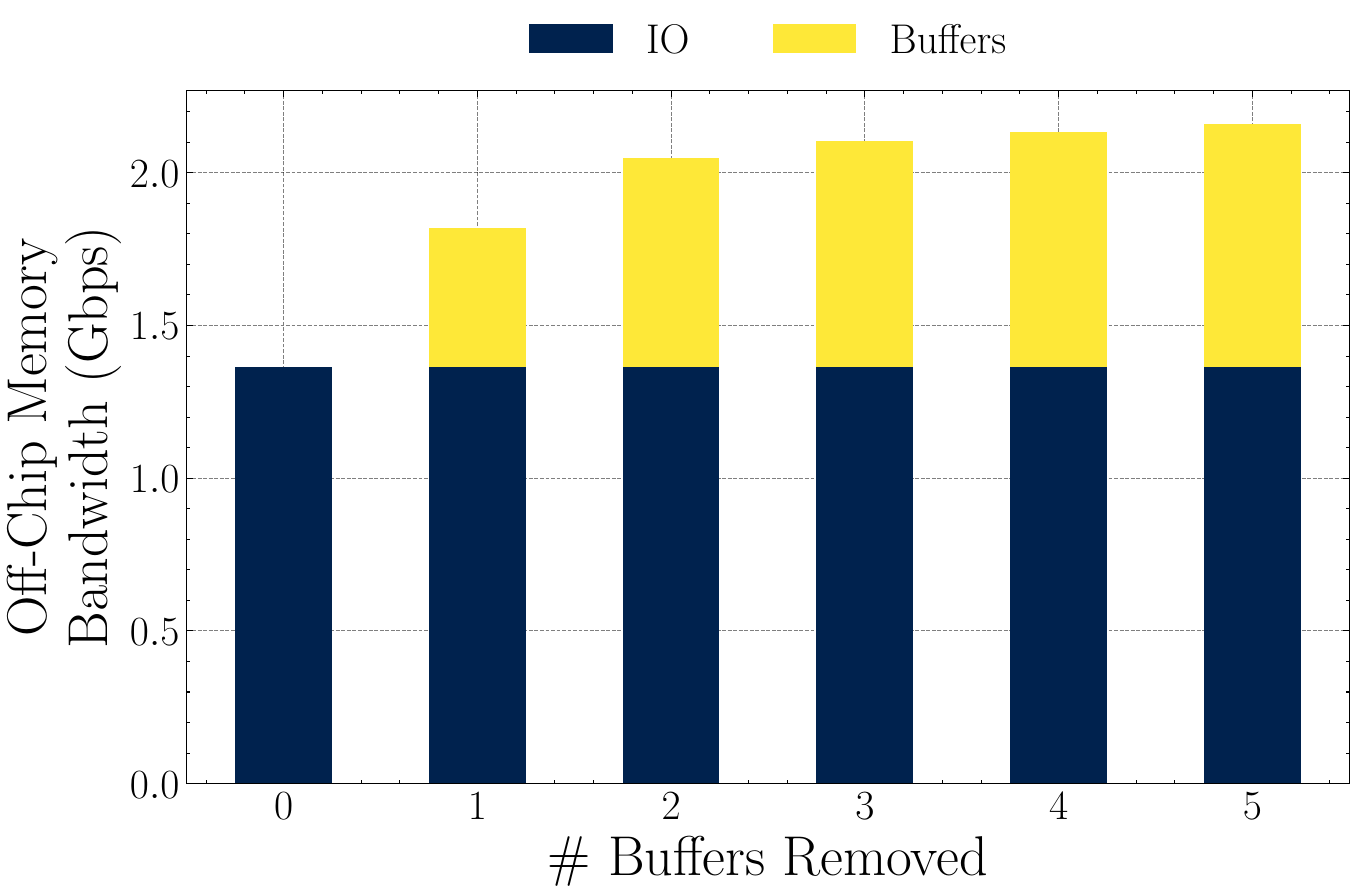}
        \caption{Off-Chip Memory Bandwidth Utilisation}
        \label{fig:ablation-buffer-bw}
    \end{subfigure}
    \caption{
        Ablation study on moving the top 5 largest buffers to off-chip memory,
        within a generated design for a 640x640 YOLOv5n model. 
        The impact on on-chip memory resource usage, off-chip memory bandwidth utilisation, and LUTRAM usage is shown.
        In Figure~\ref{fig:ablation-lutram}, the \textcolor{red}{red line} depicts the maximum LUTRAM usage for the ZCU104 FPGA platform.
    }
    \label{fig:ablation-study}
    \vspace{-0.3cm}
\end{figure*}

The software implementation in Listing~\ref{lst:soft-fifo} describes a FIFO which concurrently reads and writes to a block of memory in a first-in, first-out order.
To make the software FIFO hardware friendly, chunks of words rather than individual words are transferred, making more efficient use of the bursting abilities of the DMA. 
A choice of chunk size greater than the DMA burst size leads to no performance degradation. 

The choice of implementing buffers off-chip has implications in terms of on-chip memory resources and off-chip memory bandwidth.
To navigate this design space, we have created an automated methodology for allocating off-chip memory buffers.
To do so, we introduce the optimisation variable $t_{n,m}^{buf} \in \{ \text{ON}, \text{OFF} \}$ which dictates the location for a buffer connecting node $n$ to node $m$.

\begin{algorithm}[!t]
\caption{Buffer Allocation for Skip Connections}
\begin{algorithmic}
\LineComment{Initialise all buffers on-chip}
\State $t_{n,m}^{buf}$ = ON $\forall \; n \in [1..N], m \in [1..N]$;
\LineComment{Iterate over buffers, sorted by largest first}
\For {$n,m$ \textbf{in} sorted$\left([[1..N], [1..N]], \, d(i,j)\right)$}
    \LineComment{Check if on-chip memory is over utilised}
    \If{ $\sum_{n=1}^N\sum_{m=1}^N s_{buf}(n,m,\bm{t}^{buf}) > \mathcal{S}^{avail}$}
        \State $t_{n,m}^{buf}$ = OFF
        \Comment{Re-assign memory to off-chip}
    \Else
        \State \textbf{break}
        \Comment{Allocation Complete}        
    \EndIf
\EndFor
\end{algorithmic}
\label{alg:buffer-allocation}
\end{algorithm}

The off-chip memory bandwidth for a buffer is given as,
\begin{equation}
    \label{eq:buffer-bw}
    b_{buf}(n,m, \bm{t}^{buf}) = \begin{cases}
        2 \cdot \frac{S_{n,m} \cdot w_a}{\mathcal{L}}  & \text{if } t_{n,m}^{buf} = \text{ OFF} \\
        \hfil 0                             & \text{otherwise}
    \end{cases}
\end{equation}

where $S_{n,m} = H_{n,m} \times W_{n,m} \times C_{n,m}$ is the size of the feature map transmitted through the buffer, $w_a$ is the wordlength of the activations, and $\mathcal{L}$ is the latency of the accelerator design.
The on-chip memory requirements are described as,
\begin{equation}
    \label{eq:buffer-size}
    s_{buf}(n,m, \bm{t}^{buf}) = \begin{cases}
        q(n,m) \cdot w_a & \text{if } t_{n,m}^{buf} = \text{ ON} \\
        \hfil 0          & \text{otherwise}
    \end{cases}
\end{equation}

where $q(n,m)$ is the depth of buffer required between nodes $n$ and $m$. 
These depths are obtained from buffer depth analysis during simulation.

The objective of the buffer allocation is to discover a design which fits within on-chip memory constraints, whilst minimising off-chip memory bandwidth consumption.
Additionally, the number of off-chip memory buffers should be reduced due to the additional resource overheads of DMAs.
This optimisation objective for this is described as,
\begin{align*}
    \min_{\bm{t}^{buf}} \sum_{n=1}^N\sum_{m=1}^N b_{buf}(n,m, \bm{t}^{buf}) + 
    \lambda \cdot \sum_{n=1}^N\sum_{m=1}^N (t_{n,m}^{buf} = \text{OFF}) \\
    \text{s.t. }\sum_{n=1}^N\sum_{m=1}^N s_{buf}(n,m, \bm{t}^{buf}) < \mathcal{S}^{avail} \hfil
\end{align*}

where $\lambda$ is a regularisation hyper-parameter for penalising the number of off-chip memory buffers, and $\mathcal{S}^{avail}$ is the remaining available on-chip memory space after weights and sliding windows are allocated. 

For the streaming architecture designed in this work, the memory bandwidth overheads are insignificant (see Figure~\ref{fig:ablation-buffer-bw}).
Therefore, the buffer allocation methodology for our architecture focuses on moving the largest buffers off-chip first.
This is outlined in Algorithm~\ref{alg:buffer-allocation}.

\section{Ablation Study}

In this section, the efficacy of the proposed off-chip FIFO allocation method for supporting YOLO models within streaming architectures is evaluated. 
An ablation study was performed on the YOLOv5n model targeting the ZCU104 FPGA board. 
An accelerator architecture was generated for the given model and platform, allocating all the buffers on-chip.
The on-chip buffers were ordered by their depth, and the top five buffers were sequentially moved to off-chip memory using the software FIFO implementation. 

The results of the ablation study are presented in Figure~\ref{fig:ablation-study}.
Across the figures, it can be observed that the first three results have the greatest impact on on-chip memory resources as well as off-chip memory bandwidth utilisation.
This coincides with the buffer allocation chosen by the method described in Algorithm~\ref{alg:buffer-allocation}, demonstrating the method's effectiveness.
The impact on on-chip memory resources by re-allocating buffers can be seen in Figure~\ref{fig:ablation-buffer-size}.
Removing the first five buffers reduces the buffer memory by 56\%, with an overall reduction of 17\% for the whole on-chip memory.
For resource-constrained devices, these savings can enable the mapping of larger networks onto these devices.
The impact of this can be seen in greater detail in Figure~\ref{fig:ablation-lutram}.
The resource savings of buffer allocation can bring a design within resource constraints, as can be seen by allocating the largest buffer off-chip.

The trade-off when allocating buffers off-chip is increased memory bandwidth utilisation, which is highlighted in Figure~\ref{fig:ablation-buffer-bw}.
Moving five of the buffers off-chip leads to a 35\% increase in off-chip memory bandwidth utilisation, resulting in 2.15 Gbps of bandwidth.
However, the total off-chip memory bandwidth is insignificant compared to the 135 Gbps of off-chip memory bandwidth available.
This is one of the major benefits of streaming architectures, which make efficient use of the available on-chip memory and bandwidth.

\begin{figure}[!b]
    \centering
    \includegraphics[width=0.95\columnwidth]{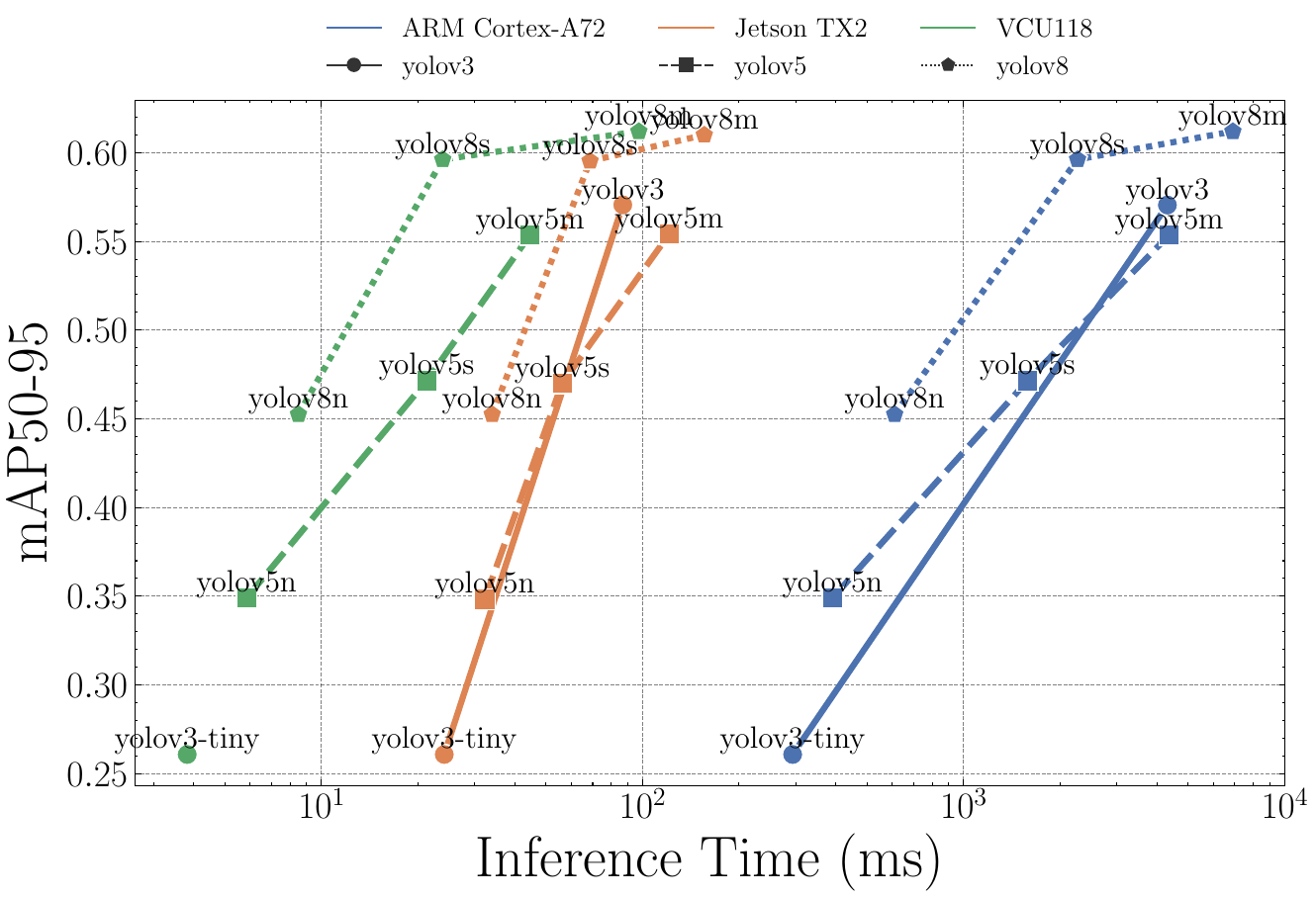}
    \caption{
        Comparison of the different versions of YOLO models (v3, v5, v8) across three different devices (\textit{ARM Cortex-A72}, \textit{Jetson TX2}, \textit{VCU118 (SATAY)}).
        The input image size used for v3 models is $416\times416$ and for the v5, v8 models is $640\times640$. 
        The results were obtained on the COCO-128 dataset.
    }
    \label{fig:yolo_comparison}
\end{figure}

\begin{table*}[htb]
\centering
\caption{
Comparison of FPGA accelerator designs generated by our toolflow (SATAY) and prior state-of-the-art works. We demonstrate much greater performance and versatility compared to previous designs.
}
\label{tab:fpga-comparison}
\resizebox{\textwidth}{!}{%
\begin{threeparttable}
\begin{tabular}{@{}rccccccc|cccccc@{}}

\toprule
                &  D. T. Nguyern \cite{nguyen2020layer} & Z. Yu \cite{yu2020} & V. Herrman \cite{herrmann2022yolo} & \multicolumn{2}{c}{D. Pestana \cite{pestana2021}} & P. Li \cite{li2021mapping} & M. Liu \cite{liu2021efficient} & \multicolumn{6}{c}{Ours}\\
\midrule
\midrule
Model                       & Sim-YOLOv2-Tiny  			& YOLOv3-Tiny & YOLOv3-Tiny	& YOLOv3-Tiny	& YOLOv4-Tiny	& YOLOv4-Tiny	& TT-YOLOv5s	& \multicolumn{2}{c}{YOLOv3-Tiny}   & \multicolumn{2}{c}{YOLOv5s}   & \multicolumn{2}{c}{YOLOv8s}\\ 
Image Size                  & 416      					& 416         & 416			& 416			& 416			& 416			& 640			& \multicolumn{2}{c}{416}    		& \multicolumn{2}{c}{640}     	& \multicolumn{2}{c}{640}\\ 
mAP\tnote{$\ast$} (\%)      & 71.13\tnote{$\dagger$}	& 30.9        & 33.1		& 33.1			& 40.2			& 40.2			& 54.6			& \multicolumn{2}{c}{33.9}   		& \multicolumn{2}{c}{56.2}    	& \multicolumn{2}{c}{61}\\ 
GFLOPs\tnote{$\ddagger$}     & 17.18    				& 5.6         & 5.6			& 5.6			& 7.5			& 7.5			& 18.9			& \multicolumn{2}{c}{5.99}  		& \multicolumn{2}{c}{18.2}  	& \multicolumn{2}{c}{30.48}\\ 
Precision                   & W1A8     					& W16A16      & W32A32		& W16A16		& W16A16		& W16A16		& -				& \multicolumn{2}{c}{W8A16} 		& \multicolumn{2}{c}{W8A16}  	& \multicolumn{2}{c}{W8A16}\\ 
\midrule
Device          			& VC707       				& ZedBoard    & Arria 10	& KU040	      	& KU040			& ZedBoard		& KCU116    	& VCU110     & VCU118    & VCU110    & VCU118    & VCU110    & VCU118  \\
LUT             			& 245K      				& 26K         & 109K (ALM)	& 139K          & 181K			& 31K			& 187K         	& 127K       & 431K      & 602K      & 1170K     & 629K      & 1023K   \\
DSP             			& 829         				& 160         & 1122		& 839           & 1255			& 149			& 1351         	& 1780       & 6687      & 1794      & 5077      & 1767      & 6815    \\
BRAM (36K)           		& 622.5  				    & 92.5        & 2195 (M20K)	& 384    		& 447			& 132 		    & 235  			& 2090.5     & 2148      & 1888      & 2026      & 2782.5    & 661     \\
URAM          			    & -           				& -           & -			& -             & -				& -				& -           	& -          & 45        & -         & 33        & -         & 713     \\
Freq. (MHz)     			& 200         				& 100         & 200			& 143           & 143			& 100			& -           	& 220        & 255       & 200       & 270       & 200       & 240     \\
\midrule
Latency (ms)    			& 9.1        				& 532 		  & 27.5        & 30.9		    & 32.1			& 18025         & 553       	& 14.3       & 6.8       & 46.4      & 14.9	     & 122.8	 & 24.5    \\ 
Energy (mJ)     			& -           				& 1788		  & -         	& 119			& - 			& 42900         & 9622        	& 220        & 252       & 1150      & 910 		 & 2890		 & 1371    \\ 
Power (W)       			& -           				& 3.4		  & -         	& 3.9			& - 			& 2.4           & 17.4      	& 15.4       & 42.9      & 24.8      & 67.0 	 & 23.5		 & 57.4    \\ 
GOP/s           			& 1877        				& 10.5  	  & 202       	& 181.4  	    & 233.6 		& 0.4           & 34.2      	& 418.9      & 875.7     & 392.0     & 1219.8	 & 248.2	 & 1244    \\ 
GOP/s/DSP       			& 2.26        				& 0.07 		  & 0.18        & 0.22 	    	& 0.19  		& 0.003         & 0.03       	& 0.24       & 0.13      & 0.22      & 0.24    	 & 0.14		 & 0.18    \\ 

\bottomrule
\end{tabular}
\begin{tablenotes}
	\item[$\ast$] mAP is reported at 0.5 IoU threshold on COCO 2017 Val dataset
        \item[$\dagger$] mAP is reported for the PASCAL VOC dataset
	\item[$\ddagger$] FLOPs are reported as MAC operations.
\end{tablenotes}
\end{threeparttable}
}
\end{table*}

\section{Evaluation}
\label{sec:eval}
In this section, we evaluate the accelerators auto-generated by SATAY. For hardware synthesis, Vivado 2020.1 is used. Our toolflow currently supports AMD FPGA devices, and we have investigated designs for the U250, ZCU104, VCU110, VCU118 devices in this evaluation.

\subsection{Comparison with other FPGA designs}

Table~\ref{tab:fpga-comparison} shows a comparison between the accelerator designs generated by our toolflow, and designs from prior works, for a selection of YOLO models and FPGA platforms.
One key benefit of our toolflow is its versatility, as we are able to produce a range of accelerators for different models of the YOLO families, including the state-of-the-art YOLOv8 models, of which this is the first FPGA implementation within the literature.
Furthermore, the generated designs achieve exceptionally high performance density, with the YOLOv8s VCU110 accelerator achieving 8\% greater GOP/s/DSP than the next best fixed-point accelerator.

The performance and accuracy of the various accelerators are explored in more detail in both Table~\ref{tab:fpga-comparison} and Figure~\ref{fig:acc-perf-pareto}. When comparing the designs with the same model, our designs achieve 1.1-3.4$\times$ GOP/s/DSP on YOLOv3-tiny compared with \cite{yu2020, herrmann2022yolo, pestana2021}, and 7.3$\times$ GOP/s/DSP on YOLOv5s compared with \cite{liu2021efficient}. 
Nguyern's design \cite{nguyen2020layer} achieved significantly higher efficiency than other related works, due to the usage of weight binarization. However, their method has only been tested on the YOLOv2 model so far. As future work, We will investigate if weight binarization can also be applied to the latest, compact YOLO-v8 model. 

Overall, our toolflow is able to generate all the points on the performance density and accuracy Pareto front, as shown in Figure~\ref{fig:acc-perf-pareto}, demonstrating both the efficiency of the streaming architecture design and the effectiveness of the design space exploration.

\begin{table}[!h]
    \centering
    \caption{Accuracy, performance, energy and power results on YOLOv5n over different FPGA platforms, as well as a Jetson TX2 GPU platform.}
    \label{tab:yolov5n_energy}
    \resizebox{0.8\columnwidth}{!}{%
    \begin{tabular}{@{}lcccc|c@{}}
    \toprule
                 & U250   & ZCU104 & VCU110 & VCU118 & Jetson TX2 \\
    \midrule
                 & \multicolumn{5}{c}{{\textbf{YOLOv5n ($320 \times 320$)}}} \\ \cline{2-6} 
    mAP50 (\%)   & \multicolumn{4}{c|}{21.2}         & 22           \\
    Latency (ms) & 3.72   & 9.83   & 4.92   & 2.21   & 10.73        \\
    Power (W)    & 115.94 & 14.82  & 23.88  & 63.27  & 6.59         \\
    Energy (mJ)  & 431.05 & 142.6  & 117.5  & 139.81 & 70.71        \\
    \midrule
                 & \multicolumn{5}{c}{{\textbf{YOLOv5n ($640 \times 640$)}}} \\ \cline{2-6} 
    mAP50 (\%)   & \multicolumn{4}{c|}{26.7}         & 27.7         \\
    Latency (ms) & 5.22   & 21.41  & 11.73  & 4.64   & 32.28        \\
    Power (W)    & 105.51 & 14.5   & 22.75  & 60.27  & 8.58         \\
    Energy (mJ)  & 550.86 & 317.33 & 266.88 & 279.8  & 276.96       \\
    \bottomrule
    \end{tabular}%
    }
\end{table}

\begin{figure}[!h]
    \centering
    \includegraphics[width=0.9\columnwidth]{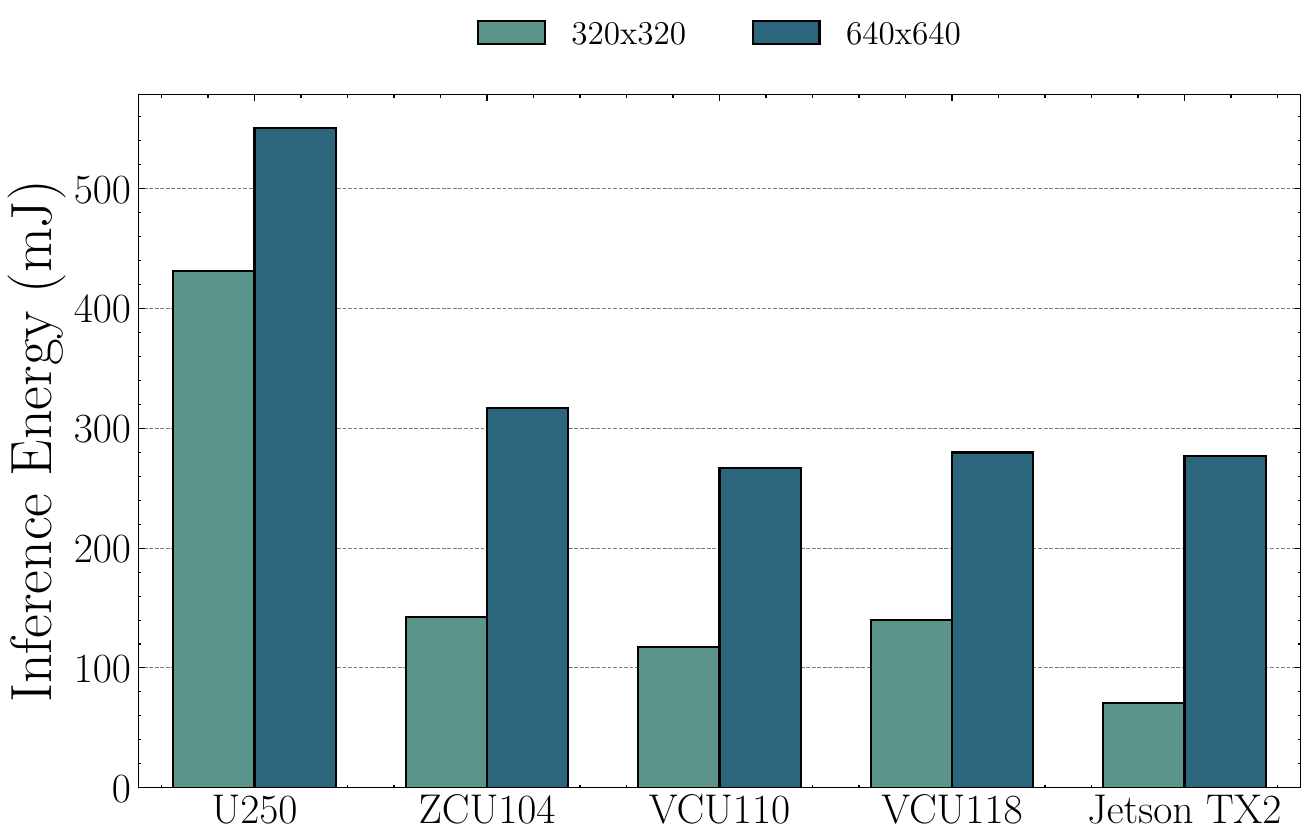}
    \caption{
        Energy per inference used during the execution of YOLOv5n models on varying image sizes ($320\times320$, $640\times640$). We compare our generated accelerator designs on different FPGA platforms to a Jetson TX2 GPU device.
    }
    \label{fig:energy}
\end{figure}

\subsection{Comparison with CPU and GPU}

To further evaluate the efficiency of our toolflow, we compare our generated FPGA designs with embedded CPU (ARM Cortex-A72) and GPU (Jetson TX2) devices in Figure~\ref{fig:yolo_comparison}. For fair comparison, we have picked FPGA designs with comparable energy consumption to the CPU and GPU, and we observe our FPGA designs have an average of 79$\times$ and 3.6$\times$ speed-up than the CPU and the GPU, over a wide range of YOLO models.  

A more detailed comparison between our auto-generated FPGA designs and the Jetson TX2 GPU can be found in Figure~\ref{fig:energy} and Table~\ref{tab:yolov5n_energy}. In Figure~\ref{fig:energy}, while the large U250 device has a much larger energy overhead, the rest of the FPGA devices have comparable energy per inference to that of the GPU, especially when accelerating the detection for the larger image size ($640\times640$). In Table~\ref{tab:yolov5n_energy}, it can be seen that the comparable ZCU104 device has much greater performance compared to the Jetson, despite similar power profiles.

\section{Conclusion}
This paper presents SATAY, a Streaming Architecture Toolflow for Accelerating YOLO networks on FPGA devices. By addressing the challenges of deploying state-of-the-art object detection models onto FPGA devices for ultra-low latency applications, SATAY generates high performance accelerators in an automatic way. Resulting accelerator designs outperform existing FPGA accelerators, demonstrating competitive performance and energy efficiency compared to CPU and GPU devices.   
\bibliographystyle{IEEEtran}
\bibliography{references}

\end{document}